\documentclass[twocolumn,showpacs,amsmath,amssymb,superscriptaddress]{revtex4}
\usepackage{graphicx}
\usepackage{dcolumn}
\usepackage{bm}
\preprint{APS/123-QED}
\begin{document}

\title{Theory of bright-state stimulated Raman adiabatic passage}
\author{G. G. Grigoryan}
\email{gaygrig@gmail.com} \affiliation{Institute for Physical
Research, 0203, Ashtarak-2, Armenia}
 \author{G. V. Nikoghosyan}
 \affiliation{Fachbereich Physik, Universit\"at Kaiserslautern, Erwin-Schr\"odinger-Strasse, D-67663
 Kaiserslautern, Germany}
\affiliation{Institute for Physical Research, 0203, Ashtarak-2,
Armenia}
 \author{T. Halfmann}
\affiliation{Institute of Applied Physics, Technical University of
Darmstadt, 64289 Darmstadt, Germany }
\author{Y. T. Pashayan-Leroy}
 \affiliation{Institut Carnot de
Bourgogne, UMR 5209 CNRS - Universit\'e de Bourgogne, BP 47870,
21078 Dijon, France}
\author{C. Leroy}
\affiliation{Institut Carnot de
Bourgogne, UMR 5209 CNRS - Universit\'e de Bourgogne, BP 47870,
21078 Dijon, France}
\author{S.~Gu\'erin}
\email{sguerin@u-bourgogne.fr}
 \affiliation{Institut Carnot de
Bourgogne, UMR 5209 CNRS - Universit\'e de Bourgogne, BP 47870,
21078 Dijon, France} \email{sguerin@u-bourgogne.fr}

\begin{abstract}
We describe analytically and numerically the process of population
transfer by stimulated Raman adiabatic passage through a bright
state when the pulses propagate in a medium. Limitations of the
adiabaticity are analyzed and interpreted in terms of reshaping of
the pulses. We find parameters for the pulses for which the
population transfer is nearly complete over long distances.
\end{abstract}

 \maketitle

\section{Introduction}
A very popular technique for complete population transfer in
lambda systems uses stimulated Raman processes by adiabatic
passage (STIRAP) \cite{STIRAP1,STIRAP2,STIRAP3,STIRAP4}. The
method is based on a so called counterintuitive sequence of laser
pulses in which the Stokes pulse, coupling the final state and the
intermediate excited state, precedes the pump, coupling the
initial state and the excited state.
The propagation of a counterintuitive pulse sequence in a
$\Lambda$ system in the STIRAP regime has been theoretically
investigated in \cite{STIRAPmedium}. An efficient population
transfer has been reported in crystals doped with rare-earth
elements Pr$^{3+}$:Y$_2$SiO$_5$ \cite{Goto,Goto2,Exp}. We note
that, due to their large density and scalability, solid media are
of significant interest for applications, e.g. in optical data
storage and processing. Particular solids, e.g. quantum dots,
colour centers or doped solids combine the advantages of atoms in
the gas phase (i.e. spectrally narrow transitions) and solids
(density and scalability). Reference \cite{Exp} also shows an
alternative efficient method of population transfer where an
intuitive sequence of pulses is used with a large one-photon
detuning. Unlike the STIRAP for which the dynamics adiabatically
projects along a dark state (i.e. no component of the excited
state), the intuitive process follows a bright state. It has been
named bright STIRAP (b-STIRAP). The possibility of such a transfer
for one atom was predicted and analyzed in
\cite{Vitanov,Topology}.

In this paper we investigate theoretically population transfer by
b-STIRAP in a medium.  A detailed study of the system of Maxwell
and Schr\"odinger equations is performed.  We derive a
self-consistent solution of the problem taking into account the
first order nonadiabatic corrections. From the solution obtained
 we derive the conditions of complete
population transfer by adiabatic passage in a medium at large
propagation distances. We show that during propagation both pulses
experience a reshaping. From the obtained solution we derive a
criterion for adiabaticity condition.

The paper is organized as follows. In Sec. II we explain a
theoretical model for b-STIRAP. The solutions to the propagation
equations are presented in Sec. III. In Sec. IV we apply the
theoretical model to a real experiment and  discuss the results
obtained comparing analytical and numerical solutions. Section V
is devoted to the pulse dynamics. The conclusion is presented in
Sec. VI.

\section{The model}
We consider the propagation of near-resonant pump and Stokes
pulses (of respective frequency $\omega_p$ and $\omega_s$) in a
medium of $\Lambda$-type atoms of ground (resp. upper) states
$|1\rangle$, and $|3\rangle$ (resp. $|2\rangle$) with the
respective energies $\omega_1$, $\omega_3$ and $\omega_2$. (see
Fig. 1). The pulses are delayed at the entrance of the medium such
that the pump pulse is switched on first. This pulse sequence is
referred to as intuitive sequence with respect to the well-known
counterintuitive order in stimulated Raman adiabatic passage
(STIRAP) processes \cite{STIRAP1}. We assume pulse durations to be
much shorter than the relaxation times, such that losses from the
upper state, and the decoherence between the two ground states due
to collisions and laser phase fluctuations are negligible. The
effect of the dissipation is considered in Section IV.C.
\begin{figure}[hbtp]
\label{Schema}
 \includegraphics[scale=0.9] {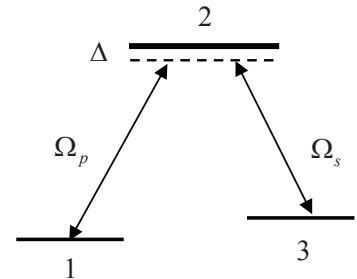}
 \caption{ \label{Scheme}
 Schematic diagram of the three-level atomic system.
 }
 \end{figure}
For an exact two-photon resonance, the corresponding Hamiltonian
in the resonant approximation (RWA) reads in the basis
$\{|1\rangle,|2\rangle,|3\rangle\}$
\begin{equation}
\label{hamilt} H =
\left( \begin{array}{ccc} {0} & {-\Omega _{p}^{*} } & {0} \\
{-\Omega _{p} } & {\Delta } & {-\Omega _{s} } \\ {0} & {-\Omega
_{s}^{* } } & 0
\end{array}
\right),
\end{equation}
with $\Omega _{p,s} =\mu _{p,s} E_{p,s} /2\hbar\equiv |\Omega
_{p,s}|e^{i\varphi_{p,s}}$ the Rabi frequencies of the pump and
Stokes laser fields, $\mu _{p,s}$ the corresponding dipole
moments, $\Delta =\omega_{2}-\omega_1 -\omega _{p} $ the
one-photon detuning. The instantaneous eigenstates write
\begin{subequations}
\label{eig_states}
\begin{eqnarray}
|b_{1}\rangle &=& \cos\psi \sin\theta e^{-i\varphi_{p}}|1\rangle +
\cos\psi \cos\theta e^{-i\varphi_{s}}|3\rangle \nonumber\\
&&+\sin\psi|2\rangle,
\\
|b_{2}\rangle &=& \sin\psi \sin\theta e^{-i\varphi_{p}}|1\rangle
+ \sin\psi \cos\theta e^{-i\varphi_{s}}|3\rangle\nonumber\\
&&-\cos\psi|2\rangle,\\
|d\rangle &=& \cos\theta e^{-i\varphi_{p}}|1\rangle - \sin\theta
e^{-i\varphi_{s}}|3\rangle
\end{eqnarray}
\end{subequations}
with $\Omega ^{2} =\left|\Omega _{p} \right|^{2} +\left|\Omega
_{s} \right|^{2} $ the generalized Rabi frequency, and the mixing
angles defined as $\tan\theta =\left|\Omega _{p} /\Omega _{s}
\right|$, $0\le\theta<\pi/2$ and $\tan2\psi = 2\Omega /\Delta $,
$0\le\psi<\pi/2$ ($0\le\psi<\pi/4$ if $\Delta>0$). Note that,
although for one atom the Rabi frequencies can be taken as real,
their phases change in general during the propagation and become
complex. The corresponding eigenvalues read

\begin{subequations}
\label{eig_values}
\begin{eqnarray}
\lambda_{b_1}&=&\frac{1}{2}\left(\Delta-\sqrt{4\Omega^2+\Delta^2}\right)=-\Omega\tan\psi,\\
\lambda_{b_2}&=&\frac{1}{2}\left(\Delta+\sqrt{4\Omega^2+\Delta^2}\right)=\Omega\,\text{cotan}\psi,\\
\lambda_{d} &=& 0.
\end{eqnarray}
\end{subequations}

State $|1\rangle$ is connected to the bright state $|b_1\rangle$
($|b_2\rangle$) when $\psi \to 0$ ($\psi \to \pi/2$),
corresponding to $\Delta>0$ ($\Delta<0$), and $\theta \to \pi /2$
which is satisfied at early times when the pump pulse is switched
on first. We will consider without loss of generality $\Delta>0$.
The bright state $|b_1\rangle$ is up to a phase the state solution
during the interaction if the adiabatic conditions are fulfilled:
\begin{subequations}
\label{adiab0}
\begin{eqnarray}
\left|\lambda_{b_1} - \lambda_{d}\right|&\gg&
|[i\dot\theta-(\dot\varphi_p-\dot\varphi_s)\cos\theta\sin\theta]\cos\psi|,\\
\left|\lambda_{b_1} -\lambda_{b_2} \right|  &\gg&
\Bigl|i\dot\psi+\frac{1}{2}(\dot\varphi_p\sin^2\theta
+\dot\varphi_s\cos^2\theta)\sin2\psi\Bigr|,\qquad
\end{eqnarray}
\end{subequations}
where the dot corresponds to the derivative with respect to time.
This is satisfied for
\begin{equation}
\label{adiab} \left|\Delta T\right|\gg1,\quad \Omega^2
T/\left|\Delta \right| \sim \left|\Delta T \right| \psi ^{2} \gg1
\end{equation}
with $T$ the time of interaction and $\psi$, $\Omega$ considered
during the pulse overlapping. The first inequality means that the
spectral width of the pulses should be much smaller than the
one-photon detuning, and the second one that this width
 should be much smaller than the Stark shift
of the levels. We have obtained the latter inequality considering
the case of interest $\Omega\le|\Delta|$, which leads to
$\sin\psi\sim\psi$ and $\cos\psi\sim1$. Note that the population
of the upper state is proportional to $\psi^{2}$ and in order to
reduce the losses from this level (of rate $\Gamma$) one should
require additionally
\begin{equation} \label{adiab_loss} \Gamma T\psi ^{2}\ll1.
\end{equation}
The Maxwell equations in the running coordinates
\begin{equation}
\eta=x,\; \tau=t-x/c
\end{equation}
and in the slowly varying amplitudes approximation read
\begin{equation}
\label{Maxwell} \frac{\partial \Omega _{p} }{\partial \eta}
=iq_{p} a_{1}^{*} a_{2} ,\quad \frac{\partial \Omega _{s}
}{\partial \eta} =iq_{s} a_{3}^{*} a_{2}
\end{equation}
where $q_{p,s} =2\pi \omega _{p,s} \mu _{p,s}^{2} N/\hbar c$ are
the coupling coefficients, with $N$ the atom density in the
medium, and $a_{i}$ the atomic population amplitudes determined
from the Schr\"odinger equation
\begin{equation}
\label{Shroedinger} i\frac{\partial}{\partial \tau} \phi=H{\rm \;
}\phi
\end{equation}
with $\phi\equiv [a_1\ a_2\ a_3]^{t}$ ($t$ denotes the transpose).
For simplicity, we consider in what follows equal oscillator
strengths $q_{p} =q_{s}\equiv q$.

\section{Effective propagation equations}
\subsection{First order equations}
Combining the Schr\"odinger equation \eqref{Shroedinger} and the
propagation equations \eqref{Maxwell} leads to
\begin{equation}
\label{Maxw2}
\frac{\partial \Omega ^{2} }{\partial \eta}
=-q\frac{\partial }{\partial \tau} \left|a_{2} \right|^{2},\quad
\frac{\partial \Omega _{s}^{*} \Omega _{p} }{\partial \eta}
=q\frac{\partial a_{3} a_{1}^{*} }{\partial \tau}  .
\end{equation}
Using the adiabatic solution of the Schr\"odinger equation, we get
finally the following effective propagation equations for the
angles $\theta$, $\psi$ and the relative phase $\varphi=\varphi
_{p} -\varphi _{s}$:
\begin{subequations}
\label{system}
\begin{eqnarray}
\label{systema} \frac{\partial \psi }{\partial \eta}
+\frac{q}{\Delta ^{2} }
\cos ^{3} 2\psi \frac{\partial \psi }{\partial \tau} &=&0, \\
\label{systemb}\frac{\partial \theta }{\partial \eta}
-\frac{q}{\Omega ^{2} }\cos ^{2}\psi
\frac{\partial \theta }{\partial \tau} &=&0, \\
\label{systemc}\frac{\partial \varphi }{\partial \eta}
-\frac{q}{\Omega ^{2} }\cos ^{2}\psi \frac{\partial
\varphi}{\partial \tau} &=&0.
\end{eqnarray}
\end{subequations}
This system of equations is valid only when the derivatives at all
orders of $\psi$, $\theta$ and $\varphi$ exist and are small such
that the adiabatic approximation is satisfied. This excludes in
particular field envelopes of bounded domains for which there
exist discontinuities of the derivative at a certain order.

We remark that this system of equations \eqref{system} contains
the time derivative of $\theta$ and $\psi$, and can be thus
interpreted as taking into account the first non-adiabatic
corrections. We indeed recover these equations starting from the
Maxwell equations (\ref{Maxw2}) to which we insert the solution to
the Schr\"odinger equation including the first order non adiabatic
corrections, i.e. keeping linear terms in $\partial_\tau\theta$
and $\partial_\tau\psi$. This calculation can be also interpreted
as an adiabatic evolution, not in the usual adiabatic basis, but
in the first order superadiabatic basis.

\subsection{Solutions}
Analytical solutions to \eqref{system} can be found by the
standard characteristic method \cite{Courant}:
\begin{equation}
\label{hodogr} \psi (\eta,\tau)=\psi _{0} (\zeta ),\; \theta
(\eta,\tau)=\theta _{0} (\xi ),\; \varphi(\eta,\tau) =\varphi _{0}
(\xi),
\end{equation}
where $\psi _{0} (\zeta)\equiv \psi(\eta=0,\tau=\zeta)$, $\theta
_{0} (\xi)\equiv \theta(\eta=0,\tau=\xi)$ and $\varphi _{0}
(\xi)\equiv \varphi(\eta=0,\tau=\xi)$ are the boundary conditions
given at the entrance of the medium, and
$\zeta\equiv \zeta(\eta,\tau)$ and $\xi\equiv \xi(\eta,\tau) $ are
solutions of the respective equations
\begin{subequations}
\label{Def_xi_zeta}
\begin{eqnarray}
\label{Def_zeta}
\zeta &=&\tau-\eta\frac{q}{\Delta ^{2} } \cos ^{3} 2\psi _{0} (\zeta ), \\
\label{Def_xi}\int _{\zeta }^{\xi }\Omega _{0}^{2}
(t)dt&=&q\eta\cos ^{4} \psi _{0}(\zeta) (2-\cos 2\psi _{0}(\zeta))
\end{eqnarray}
\end{subequations}
with $\Omega (\eta,\tau)=\Omega (\eta=0,\tau=t)\equiv\Omega_0(t)$
the generalized Rabi frequency given at the entrance of the
medium. The pump and Stokes Rabi frequencies are respectively
determined from $\Omega_p=\Omega\sin\theta$ and
$\Omega_s=\Omega\cos\theta$.

Through the definition $\tan2\psi=2\Omega/\Delta$, Eq.
\eqref{systema} can be interpreted as describing the dynamics of
the generalized Rabi frequency (for a constant $\Delta$)
$\Omega(\eta,\tau)=\Omega_0(\zeta)$. We conclude that it
propagates in the medium through the non-linear time $\zeta$ and
with the non-linear velocity $v$ such that $1/v=1/c+q\cos ^{3}
2\psi _{0}/\Delta ^{2} $, smaller than the light velocity $c$, and
the delay in the medium for small angles $\psi$ at a certain length $L$ is equal to
\begin{equation}
\label{delay} \tau_{m}=L(1/v -1/c )=\frac{qL}{\Delta ^{2} } \cos
^{3} 2\psi _{0} \sim \frac{qL}{\Delta ^{2} }.
\end{equation}

From Eq. \eqref{Def_xi}, the non-linear time $\xi $ is always
larger than $\zeta $, meaning that the mixing angle propagates in
the medium with the velocity greater than the light velocity
\cite{Proc,Li}. This can be also inferred from Eq.
(\ref{systemb}), where the non linear velocity appears negative.

The solution \eqref{hodogr} shows that, if at the medium entrance
the relative phase $\varphi$ of the pulses is constant, it remains
constant during the propagation in the medium.

\subsection{Limitations for the adiabatic passage}
We show several limitations of solution \eqref{hodogr}, all
related to the adiabaticity of the process. They can be
interpreted in terms of energy leading to the maximum propagation
length [see Eq. \eqref{Z_max}] and to the maximum time [see Eq.
\eqref{taumax}]. Non-adiabatic phenomena that lead to a
singularity in the solution, and corresponding to a reshaping of
the pulses, are prevented through the limitation \eqref{adiab2}
[or \eqref{adiab2_} in the limit of small angle $\psi$]. Reshaping
of the pulses is also prevented by \eqref{adiab3}.

\subsubsection{Energy}
At the beginning of the interaction $\tau\rightarrow-\infty$, i.e.
$\zeta\rightarrow-\infty$ according to Eq. \eqref{Def_zeta}, the
quantity $\xi(\eta,\tau \to\ -\infty)$ is estimated by the equation
\begin{equation}
    \label{Ksi}
    \int _{-\infty}^{\xi(\eta,\tau \to\ - \infty) }\Omega _{0}^{2} (\tau)d\tau \sim q \eta.
\end{equation}
We infer that $\xi(\eta,\tau\to-\infty)$ is a monotonic increasing
function of $\eta$ changing from $-\infty$ at $\eta=0$ to
$+\infty$ at $\eta = \eta_{\max}$, where $\eta_{\max}$ is
estimated from
\begin{equation}
    \label{Z_max}
    \int _{-\infty}^{+\infty }\Omega _{0}^{2} (\tau)d\tau \sim q \eta_{\max}.
\end{equation}
This gives a relation between the energy of the pulses and the
maximum theoretical propagation length. The physical
interpretation of the condition (\ref{Z_max}) for the length can
be expressed in terms of energy: the number of atoms whose
population can be fully transferred by adiabatic passage from
state $|1\rangle$ to state $|3\rangle$ in the medium cannot exceed
the number of photons in the pump and Stokes fields.

\subsubsection{Time}
The definition of $\zeta$ \eqref{Def_zeta} shows an additional
limitation: For a given length $\eta$, one can get a large value
for $\zeta(\eta,\tau)$ when taking a large $\tau$, i.e. when
$\tau\gg q\eta/\Delta^2$.
The definition of $\xi$ \eqref{Def_xi} shows then that for a too
large $\zeta$, $\xi$ does not exist. For a given $\eta$, the
maximum value $\xi\to+\infty$ is obtained for $\tau_{\max}$
estimated from
\begin{equation}
\label{taumax} \int _{\tau_{\max}}^{+\infty }\Omega _{0}^{2}
(t)dt\sim q\eta.
\end{equation}
$\tau_{\max}$ has to be interpreted as a maximum time until which
the adiabaticity is preserved. Beyond this value of $\tau_{\max}$,
one expects non-adiabatic effects (beyond the first superadiabatic
order). Eq. \eqref{taumax} means that, during the propagation
corresponding to larger $\eta$, the maximum time from which
adiabaticity is broken becomes smaller.

\subsubsection{Reshaping of the pulses}
Adiabatic conditions (\ref{adiab}) have to be revised as follows
when considering propagation. The time derivatives in Eq.
(\ref{adiab0}) (in the running coordinates) become:
\begin{equation}
\frac{\partial\psi}{\partial\tau}=\frac{d\psi_0}{d\zeta}
\frac{\partial\zeta}{\partial\tau},\quad
\frac{\partial\theta}{\partial\tau}=\frac{d\theta_0}{d\xi}
\frac{\partial\xi}{\partial\tau},\quad
\frac{\partial\varphi}{\partial\tau}=\frac{d\varphi_0}{d\xi}
\frac{\partial\xi}{\partial\tau}.
\end{equation}
This shows that the adiabatic conditions are the ones at the
entrance of the medium provided that
\begin{equation}
\frac{\partial\zeta}{\partial\tau}\le1,\quad
\frac{\partial\xi}{\partial\tau}\le1.
\end{equation}
The breaking of these conditions corresponds generally to the
formation of shock-waves during the propagation. We obtain for the
variable $\zeta$
\begin{equation}
\label{deriv} \frac{\partial\zeta}{\partial\tau} =
\frac{1}{1-(6q\eta/\Delta^2)(d\psi_0 /d\zeta)\cos ^{2} 2\psi_0
\sin 2\psi_0 }.
\end{equation}
At large propagation lengths the denominator in the right-hand
side of this expression can become very small, that would lead to
a strong increase of the derivative of the left-hand side. Thus,
in order that the adiabaticity be preserved
during propagation it is necessary to add to conditions
\eqref{adiab}:
\begin{equation}
6\psi _{0}\tan 2\psi _{0}\frac{\tau _{m}}{T}=6\psi _{0}\tan 2\psi _{0}\frac{%
qL}{\Delta ^{2}}\frac{\cos ^{3}2\psi _{0}}{T}\ll 1.
\label{adiab2}
\end{equation}
 We infer that for relatively small
angles $\psi $ and propagation lengths $L$ at which the group
delay of the generalized Rabi frequency is of the order of the
characteristic interaction time, the adiabaticity in the medium
does not break down. Considering for simplicity the case of small
angles $\psi_0$, we get the condition at the length $\eta =L$:
\begin{equation}
\label{adiab2_} qTL \ll \frac{(\Delta
T)^2}{12}\left(\frac{\Delta}{\Omega_{\max}}\right)^2
\end{equation}
with $\Omega_{\max}$ the maximum value of $\Omega$. At a given
peak Rabi frequency, a larger $\Delta$ allows thus one to increase
the medium length for which pulses can propagate without
reshaping. One can interpret this effect with the argument that a
larger detuning weakens the effective interaction of the pulses
with the medium.

With respect to the variable $\xi$, one obtains
\begin{equation}
\label{adiabKsi2}
 \frac{\partial \xi}{\partial\tau} = \frac{\Omega^2_0(\zeta)}{\Omega^2_0(\xi)}
 \frac{\partial \zeta}{\partial \tau}\left(1-\frac{6q\eta}{\Delta^2}\frac{d\psi_0}{d\zeta}
 \cos ^{2} 2\psi_0\sin 2\psi_0\right).
\end{equation}
Taking into account the preceding condition (\ref{adiab2}), this
leads to
\begin{equation}
\label{adiabKsi2_}
 \frac{\partial \xi}{\partial\tau} \sim\frac{\Omega^2_0(\zeta)}{\Omega^2_0(\xi)}
 \frac{\partial \zeta}{\partial \tau}.
\end{equation}
This shows that the part $\dot\theta$ (and also $\dot\varphi$) of
the nonadiabatic coupling at the entrance of the medium is
multiplied by the scale factor $\Omega^2_0(\zeta)/\Omega^2_0(\xi)$
during the propagation. The adiabaticity will break down when this
ratio is too large, i.e. when $\xi>\zeta$ for late times. In
practice, one can estimate the condition on the propagation length
$\eta=L$ by preventing this situation, imposing $\xi-\zeta\ll T$,
which gives from Eq. (\ref{Def_xi})
\begin{equation}
\label{adiab3} qT L\ll (\Omega_{\max}T)^2.
\end{equation}
Inequalities (\ref{adiab2}) and (\ref{adiab3}) guarantee the
adiabaticity conditions during the pulse propagation, preventing a
significant reshaping of the pulses.

\section{Population transfer by b-STIRAP}
We consider the process of complete population transfer from state
1 to state 3 by adiabatic passage through the bright state using a
so-called intuitive sequence of pulses, i.e. with the pump field
switched on before the Stokes field with a delay $\tau_d>0$. The
aim of this section is devoted to the theoretical interpretation
of this process. We consider pulses of Gaussian envelopes:
$\Omega_{p}=\Omega_{p,\max}\exp[-((t+\tau_d/2)/T_p)^2]$,
$\Omega_{s}=\Omega_{s,\max}\exp[-((t-\tau_d/2)/T_{s})^2]$.

The main limitation for an adiabatic transfer for a single atom is
given by (\ref{adiab}). It gives (i) lower and upper limits for
the detuning (at given fields) and (ii) a better adiabaticity for
a larger pulse overlapping (that can be achieved with larger Rabi
frequencies for given delay and pulse shapes). As studied above,
there are four additional limitations given by \eqref{Z_max},
\eqref{taumax}, (\ref{adiab2}) [or (\ref{adiab2_}) for small angle
$\psi_0$]  and (\ref{adiab3}). The energetic argument
\eqref{Z_max} confirms the need of large Rabi frequencies to
preserve the adiabaticity during the propagation. The condition
\eqref{taumax} on the other hand favours faster processes, i.e.
finishing before $\tau_{\max}$. We show below that this is the
case for larger detunings. Condition (\ref{adiab2_}) confirms that
a larger detuning allows the propagation over a longer distance.
Condition (\ref{adiab3}) is shown below to give the practical
limitation on the medium length for which nearly complete
population transfer can occur.
 Due to the
interaction with the atoms, the pulses change their shapes. This
corresponds to the violation of the adiabaticity conditions, as is
studied below.
We consider numerical calculations in different situations and
interpret them in terms of the limitations described above.

\subsection {Application in experimental conditions}
\begin{figure}[hbtp]
\label{ExpTheory}
 \includegraphics[scale=0.85] {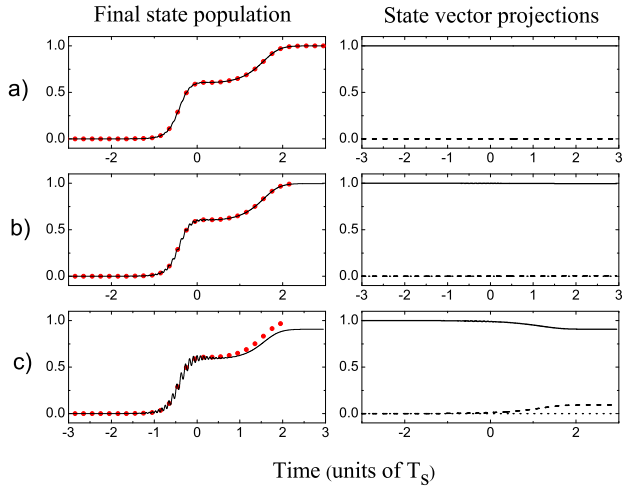}
 \caption{\label{Dyn_exp}(Color online) For (a) $qT_sx = 0$,  (b) $qT_sx = 2$, and (c)
 $qT_sx = 5$, as a function of the running time $\tau=t-x/c$, left column:
 Population transfer to state 3 determined numerically (full line)
 and from the analytical solution \eqref{hodogr} (dotted line); right column:
 Projections $|\langle d|\phi\rangle|^2$ (dashed line) and $|\langle b_2|\phi\rangle|^2$
 (dotted line).
 }
 \end{figure}

We first apply the analysis to the experiment presented in
\cite{Exp}, for which $\Omega_{p,\max}T_{s}=108.6$,
$\Omega_{s,\max}T_{s}=110.5$, $\tau_d / T_{s}=1.4$,
$T_{p}/T_{s}=0.8$, and $\Delta T_{s}=50$. Figure \ref{Dyn_exp}
displays the dynamics of population transfer to state 3 for
various lengths $qT_{s}x$ of the medium and the corresponding
projections of the state vector $| \phi \rangle$ on the
eigenstates. The analytic solution \eqref{hodogr} fits well the
numerics until the complete transfer fails. The efficiency of the
population transfer decreases as the pulses propagate into the
medium. The population transfer occurs completely in the medium up
to $qT_{s}x\approx4$. For $qT_sx=5$, one can already notice a
partial transfer of population.
\begin{figure}[hbtp]
 \centering
 \includegraphics[scale=0.9] {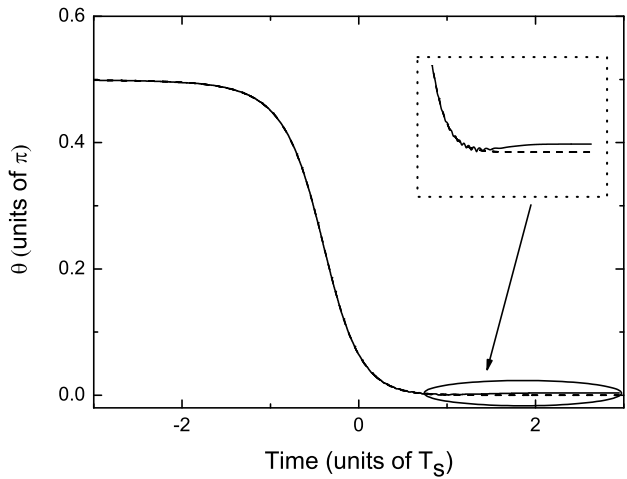}
 \caption{\label{Mix_exp}Time evolution of the mixing angle $\theta$ for
 $qT_sx = 0$ (dashed line) and  $qT_sx=5$ (solid line). The inset shows
an enlargement of the values of the mixing angle near the end of
the interaction.}
 \end{figure}
To achieve an effective population transfer one should provide an
adiabatic interaction to make state vector $| \phi \rangle$ follow
$|b_{1}\rangle$ state and a smooth change of the mixing angle
$\theta(\xi)$ from $\pi/2$ to $0$. However, during the
propagation, one notices that state vector $| \phi \rangle$
acquires a component along the dark state $|d\rangle$ leading to
some final population to state $|1\rangle$. This is clearly seen
from the right column of Fig. 2 which shows the instantaneous
projections of state vector $| \phi \rangle$ onto the dressed
states. We can see that at $qT_sx=2$, there is no projections onto
states $|b_{2}\rangle$ and $|d\rangle$, while at $qT_sx=5$ a
noticeable projection onto state $|d\rangle$ appears. As a result
the mixing angle $\theta(\xi)$
 changes and no more satisfies the conditions
that are necessary to realize the transfer process.  Indeed, in
Fig. 3 we present the time evolution of the mixing angle
$\theta(\xi)$ for the same parameters as in Fig. 2. One can see
that already at propagation length $qT_sx=5$ the final value of
$\theta(\xi)$ is different from $0$.

\subsection{Interpretation and improvement of the transfer}

\begin{figure}[h]
 \includegraphics[scale=0.7]{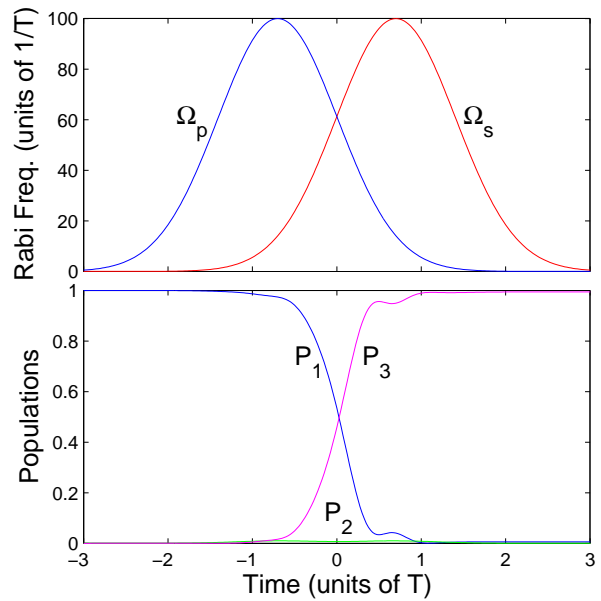}
 \caption{\label{Delta_1000} (Color online) Dynamics of the population transfer
$P_j=|\langle j|\phi\rangle|^2$ (lower frame) corresponding to the
pulses (upper frame) at the entrance of the medium $x=0$ for
$T\Delta=1000$ and $T\Omega_{p,\max}=T\Omega_{s,\max}=100$.
 }
 \end{figure}

To interpret the maximal length leading to the complete transfer
and to show how to improve it, we analyze the transfer taking
similar parameters of the experiment. We take for simplicity equal
peak Rabi frequencies and equal durations for both pulses.

\begin{figure}[h]
 \includegraphics[scale=0.7] {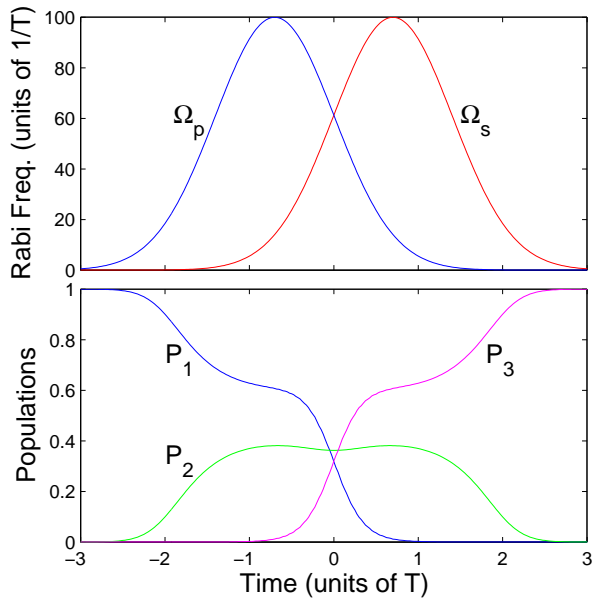}
 \caption{\label{Delta_50} (Color online) Same as Fig. \ref{Delta_1000}
but for $T\Delta=50$ and $T\Omega_{p,\max}=T\Omega_{s,\max}=100$.
 }
 \end{figure}

\begin{figure}[h]
 \includegraphics[scale=0.7] {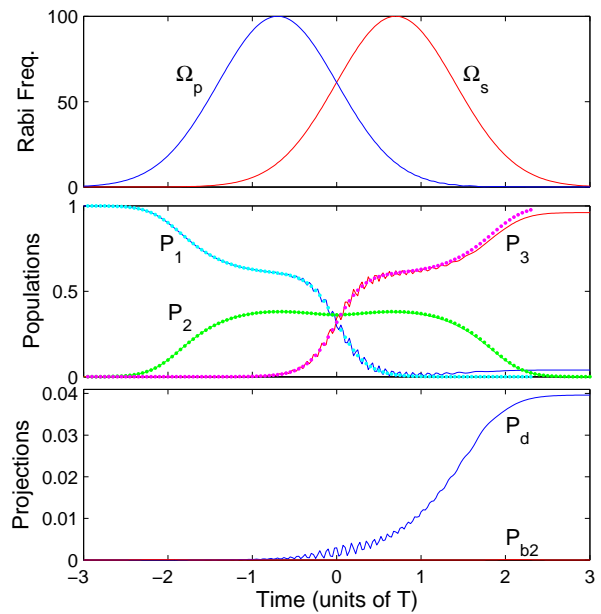}
 \caption{\label{Delta_50_z7} (Color online) Dynamics of the population transfer
$P_j=|\langle j|\phi\rangle|^2$ (middle frame) for $T\Delta=50$,
$T\Omega_{p,\max}=T\Omega_{s,\max}=100$ and at $qTx=7$
(corresponding to Fig. \ref{Delta_50} for $x=0$); the absolute
value of the corresponding Rabi frequencies (upper frame); and its
corresponding non-adiabatic projections onto the dark state
$P_d=|\langle d|\phi\rangle|^2$ and the bright state
$P_{b_2}=|\langle b_2|\phi\rangle|^2$.
 The dotted lines for the populations are determined from
 the analytic solution \eqref{hodogr}, \eqref{Def_xi_zeta}.
 }
 \end{figure}

The main limitation is here given by Eq. \eqref{taumax}. It shows
indeed that a process corresponding to a population transfer
lasting too long, i.e. beyond $\tau_{\max}$, leads to an
inefficient non-adiabatic transfer when propagation is considered.
To illustrate this effect, we consider below two limiting cases
with the same Rabi frequencies and pulse durations
($\Omega_{p,\max}=\Omega_{s,\max}$, $T_p=T_s=T$), the delay
$\tau_d=1.3 T$, and satisfying \eqref{adiab}: one with a large
detuning defined as $\Delta\gg\Omega_{\max}$ (see Fig.
\ref{Delta_1000}), leading to a negligible population in the
intermediate state,
and another one with a small detuning defined as
$\Delta\sim\Omega_{\max}$ (see Fig. \ref{Delta_50}), leading to a
noticeable population in the intermediate state. At the entrance
of the medium, $\eta=0$, the population transfer for the large
detuning case is accomplished approximately during the overlap of
the pulses (see Fig. \ref{Delta_1000}), while, for the small
detuning, it occurs for the duration of both pulses (see Fig.
\ref{Delta_50}). (The calculation with the small detuning
corresponds to conditions that are very similar to the experiment
described above.) Thus condition (\ref{taumax}) is easier to
satisfy for a larger detuning. The effect that the transient
population in the intermediate state is smaller for a larger
$\Delta$ corresponds to a weaker interaction of the pulses with
the medium, and the condition (\ref{adiab2}) [or (\ref{adiab2_})]
is also easier to satisfy. One thus expects an adiabatic transfer
over much longer propagation length for the case of large
detuning. This is confirmed by the numerics. Figure
\ref{Delta_50_z7} shows that, already for $qTx=7$, the population
transfer is not complete. One can see that the corresponding
analytic solution fits well the numerics (until it stops at
$\tau\approx2.35 T$), before the Stokes pulse is off and the
transfer is complete. This time corresponds with a quite good
approximation to the one determined from the estimation
\eqref{taumax} giving $\tau_{\max}\approx2.33 T$. Figure
\ref{Delta_50_z7} shows that the nonadiabatic corrections
correspond to a projection onto the dark state that grows roughly
monotonically as a function of time, reaching its peak value near
$\tau_{\max}$.

\begin{figure}[h]
 \includegraphics[scale=0.7] {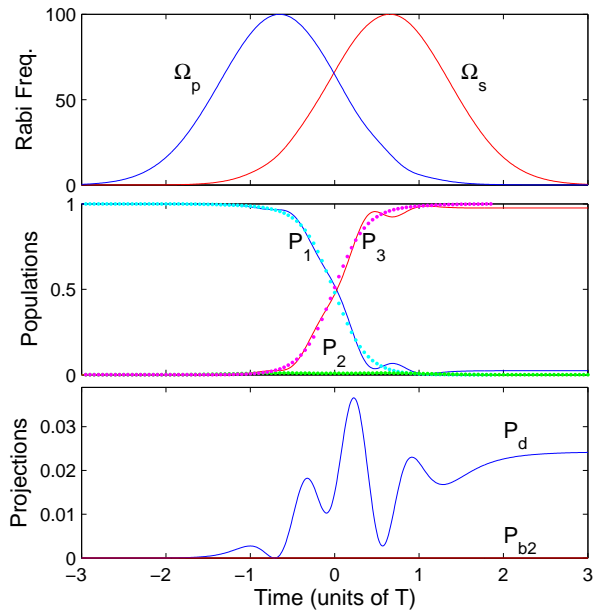}
 \caption{\label{Delta_1000_z100} (Color online) Same as Fig. \ref{Delta_50_z7}
but for $T\Delta=1000$ and $T\Omega_{p,\max}=T\Omega_{s,\max}=100$
and at $qTx=100$ (corresponding to Fig. \ref{Delta_1000} for
$x=0$).
 The small discrepancy between numerics and
 the analytic solution is due to nonadiabatic corrections of order
 higher than one.
 }
 \end{figure}

For the large detuning case, the limitation of the population
transfer during the propagation is given by \eqref{adiab3}. Figure
\ref{Delta_1000_z100} shows the dynamics for $qTx_{\max}=100$
which corresponds roughly to the maximum length with an efficient
population transfer (it reaches here approximately $99\%$). With
respect to the condition \eqref{adiab3}, this corresponds to
$qTL/(\Omega_{\max}T)^2=0.01$. Inspection of the obtained pulse
shows that this non-adiabatic effect occurs near the end of the
pump pulse, where it is significantly reshaped featuring a longer
tail. Such a reshaping breaks the initial intuitive sequence of
the pulses since when the Stokes falls down the pump pulse is not
any more completely switched off. Figure \ref{Delta_1000_z100}
shows that the nonadiabatic corrections correspond to a projection
onto the dark state. The maximal length obtained for the large
detuning case is well beyond the one obtained for the small
detuning case as discussed above.

We conclude that in order to increase the propagation length for
given identical pump and Stokes peak amplitudes, besides taking
large detuning $\Delta$ [in the limit of Eq. \eqref{adiab}], we
should take the delay between the pulses as large as possible,
while keeping an overlapping between the two Rabi frequencies of
sufficiently large area (much bigger than one).

We remark that the obtained maximum lengths of the medium where
the complete population transfer occurs are here well below the
lengths that would lead to a singularity in the pulse shapes, as
given by condition \eqref{adiab2_}.



%

\subsection{Practical implementation and examples}
Equations \eqref{adiab} and \eqref{adiab_loss} are the conditions
that must be met in order to achieve an efficient population
transfer by b-STIRAP in a medium. We consider below various media
and experimental conditions that satisfy these conditions.

In the gas phase, the limitations imposed by the losses from the
three-state system are mainly due to the inhomogeneous linewidth
of the excited state [through Eq. \eqref{adiab_loss}] and to the
ionization by the fields, mainly from the excited state. In the
case of a warm vapor, the inhomogeneous linewidth is mainly
determined by the Doppler broadening, typically $\Gamma=500$MHz
for a warm alkali-metal-atomic vapor. In a regime corresponding to
$\Omega\sim\Delta$, leading to a relatively short propagation
length and a quite large transient population in the excited state
as typically shown in Figs. 6 and 9, all the criteria are
fulfilled for a pulse duration of order $T= 40$ ps. In the
conditions of Fig. 9, we have $\Omega=\Delta/2$,
$(\Omega/\Delta)^2\Gamma T= 0.005$ (corresponding to a negligible
loss), $\Delta T=40$, $\Omega^2T/\Delta=10$ and $\Omega=500$ GHz.
This imposes to find a particular system with a limited ionization
from the excited state with such a Rabi frequency, which
corresponds to typical intensities of 30 MW/cm$^{2}$ (for a
transition strength of order of one Debye). For a long propagation
length regime (as typically shown in Fig. 7, for
$\Omega/\Delta=0.1$) with a small transient population in the
excited state, one obtains $T=1$ ns as the typical pulse duration.
In the conditions of Fig. 7, this gives $(\Omega/\Delta)^2\Gamma
T= 0.005$, $\Delta T=1000$, and $\Omega^2T/\Delta=100$
corresponding to the Rabi frequency $\Omega=100$ GHz, i.e. to a
generally weakly ionizing field of 1 MW/cm$^{2}$ (for a transition
strength of order of one Debye). We remark that such a situation
leads to $\Gamma T\sim0.5$. However it is not expected to give a
large loss due to the small transient population in the upper
state (of order $(\Omega/\Delta)^2\sim0.01$). We have checked
numerically this assumption as shown in Fig. 8, where various loss
rates are studied. The loss rate $\Gamma T=0.1$ does not show a
noticeable loss in the process, while the loss rate $\Gamma T=0.5$
gives a small loss of less than 3 percents.

\begin{figure}[h]
 \includegraphics[scale=0.9] {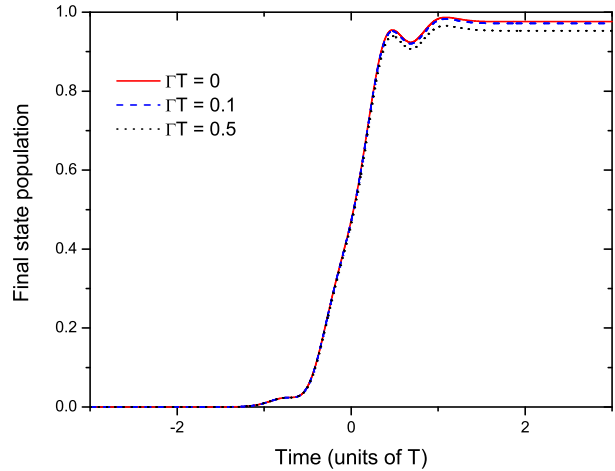}
 \caption{\label{Delta_1000_z100_Gamma} (Color online) Population transfer
 in the conditions of Fig. \ref{Delta_1000_z100} but for various
 loss rates: $\Gamma T=0$ (full line), $\Gamma T=0.1$ (dashed line,
 almost undistinguishable from the full line)
 and $\Gamma T=0.5$ (dotted line).}
 \end{figure}

The situation is less restrictive in cold medium (such as Bose
Einstein Condensates system) where $\Gamma$ is bounded by the
natural rate of losses from the excited state which is of order of
10 MHz. In this case the pulse duration can be increased to
$T\sim2$ ns for the short propagation length regime case
($\Delta\sim\Omega$) and to $T\sim50$ ns for the long propagation
length regime $\Omega/\Delta\sim0.1$. In these cases, the Rabi
frequencies range from $\Omega\sim2$ GHz to $\Omega\sim20$ GHz
which are well below the ionization regime in general.

For solids, one of the main additional difficulties comes from
their large optical inhomogeneous broadening. For example, in
rare-earth-doped crystals, that are preferable within the context
of coherent optical behavior for $f-f$ transitions, the upper
state inhomogeneous broadening is of the order of 10 GHz. However,
there are some techniques like hole-burning techniques that allow
one to select a subset of the ions within a particular spectral
range and thus to reduce the large inhomogeneous broadening. This
allows one to increase the pulse durations up to the microsecond
regime. This technique has been used in the experiment of Ref.
\cite{Exp} that we have taken to apply our theoretical results
(see Sec. IV A). There is usually another issue due to the level
splitting $\Delta_S$ in "atom-like" systems (doped solids, color
centers, quantum dots) in the range of 10 MHz - 10 GHz. This gives
limits for the Rabi frequencies to address single levels:
$\Omega\ll\Delta_S$.


\section{Pulse dynamics and reshaping}

We here show the dynamics of the intuitive pulse sequence
propagating in the medium over long distance beyond adiabatic
limitations. We choose parameters that show a reshaping already
for small distances in the medium: $\Omega_{\max}T=20$, $\tau_{d}
/ T=1$, and $\Delta T=40$.

\begin{figure}[hbtp]
 \includegraphics[scale=0.85] {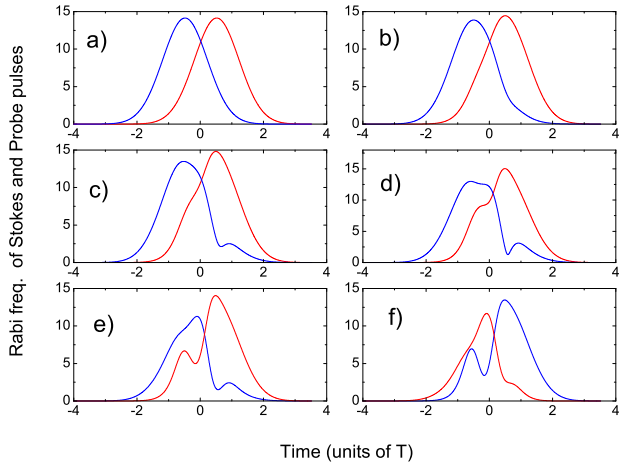}
 \caption{\label{PulseDyn} (Color online) Time evolution of the normalized
 Rabi frequencies of the probe and Stokes pulses for different propagation
 lengths $z\equiv qTx$: a) z = 0; b) z = 10; c) z = 20; d) z = 30; e) z=40;
 f) z=50.}
 \end{figure}

%

Figure \ref{PulseDyn} shows a typical dynamics: After a certain
length there occurs a noticeable reshaping corresponding to a
lengthening of the shape of the pump intensity at the tail leading
to additional peaks (here one peak is shown). Already for $z=10$,
the population transfer is not anymore complete in this situation.
The adiabaticity is strongly violated beyond this value. The
leading edge of the stokes pulse is next also reshaped also
showing one additional peak. The sequence of pulses is not any
more intuitive at large distances.


\section{Conclusion and discussion}
We have analyzed the complete population transfer by adiabatic
passage when the pulses propagate in a medium in an intuitive
sequence. Using the analytic solution, determined from the first
order nonadiabatic corrections, we have derived limitations on the
maximal length of the medium where the complete population
transfer can take place. Assuming the adiabatic conditions
fulfilled at the entrance of the medium [see conditions
\eqref{adiab}], the complete population occurs over a longer
distance in the medium (i) for a larger one-photon detuning (at
given peak Rabi frequencies of the pulses) and (ii) for a larger
peak Rabi frequency of the pulses (at a given ratio of the
one-photon detuning with the peak Rabi frequency of the pulses).
In the limit $\Delta\gg\Omega$, the limitation of the transfer
appears as non-adiabatic corrections and a reshaping of the pump
pulse through the condition \eqref{adiab3}.

We can list the differences between the techniques of STIRAP and
b-STIRAP. Unlike STIRAP, b-STIRAP depends on the one-photon
detuning. Increasing the one-photon detuning leads to an
improvement of the propagation length at which the population
transfer is complete in a medium. On the other hand, increasing
the one-photon detuning weakens the adiabaticity of the
interaction.

 The second difference concerns the derived criteria of
adiabaticity in a medium. During the propagation of the two pulses
there is an exchange of energy between them that causes their
reshaping which breaks down the adiabaticity. For a
counterintuitive pulse sequence, in the case of equal oscillation
strengths, the adiabaticity is preserved during propagation at any
propagation length. Only in the case of unequal oscillator
strengths it can break down, mainly due to a singularity
corresponding to a steepness occurring in the shape of the pulse
(see \cite{STIRAPmedium,Fleisch}). In the case of an intuitive
pulse sequence as shown in the present work the adiabaticity is
broken down in a medium even for equal oscillator strengths.

As noticed very recently \cite{Proc,Li}, an intuitive sequence of
the switching of the pulse features a reshaping that can lead to a
superluminal propagation \cite{Superl1,Superl2,Superl3,Superl4}.
Our analytic solution should provide new insights about this
effect and its limitation. Such a study is in progress.

b-STIRAP can be viewed as an alternative population transfer
method. The superluminal property of b-STIRAP can find
applications. For instance, the simultaneous use of the b-STIRAP
and STIRAP processes in multi-level systems (e.g. tripod,
M-system) can be an efficient method for creating time controlled
excitations of target states. We also anticipate that a
combination of STIRAP/b-STIRAP is of interest in quantum
information. The counterintuitive STIRAP scheme indeed allows the
population to be transferred from 1 to 3 - while the same pulse
sequence permits transfer back from 3 to 1 in an intuitive
b-STIRAP scheme. This could find applications for the
implementation of quantum gates.

\begin{acknowledgments}
We acknowledge supports from ANSEF no. PS-opt-1347, INTAS
06-1000017-9234, the Agence Nationale de la Recherche (ANR CoMoC),
the European Commission project FASTQUAST, and the Conseil
R\'{e}gional de Bourgogne. G.G. thanks the Universit\'{e} de
Bourgogne for her stay during which a part of this work was
accomplished. G.N. gratefully acknowledges the support from the
Alexander von Humboldt Foundation.
\end{acknowledgments}

\begin{appendix}
\section{Method of characteristics for $\psi$ and $\theta$}
In this appendix we derive the solution of Eqs. (\ref{system}) of
the form
\begin{subequations}
\label{system_app}
\begin{eqnarray}
\label{system_appa} \frac{\partial \psi }{\partial \eta}
+a(\psi) \frac{\partial \psi }{\partial \tau} &=&0, \\
\label{system_appb}\frac{\partial \theta }{\partial \eta} -b(\psi
) \frac{\partial \theta }{\partial \tau} &=&0,
\end{eqnarray}
\end{subequations}
using the method of characteristics \cite{Courant}. We first solve
Eq. (\ref{system_appa}) using a characteristic parametrization
$\eta\equiv\eta(s)$ and $\tau\equiv\tau(s)$ such that
\begin{subequations}
\begin{eqnarray}
\label{deta_appa} \frac{d\psi}{ds}&=&0,\\
\label{deta_appb} \frac{d\eta}{ds} &=&1, \\
\label{dtau_appc}\frac{d\tau}{ds}&=&a(\psi).
\end{eqnarray}
\end{subequations}
$\psi$ is constant along the characteristics:
$\psi(\eta,\tau)=\psi(\eta(0),\tau(0))$. From (\ref{deta_appb}),
we get $\eta(s)=s$ choosing $\eta(0)=0$. This entails
\begin{equation}
\psi(\eta,\tau)=\psi(0,\zeta)\equiv\psi_0(\zeta),\quad \zeta\equiv
\tau(0)
\end{equation}
with $\zeta\equiv\zeta(\eta,\tau)$ solution of Eq.
(\ref{dtau_appc}):
\begin{equation}
\zeta=\tau-\eta a(\psi_0(\zeta)).
\end{equation}
This leads to Eq. (\ref{Def_zeta}).

We next solve Eq. (\ref{system_appb}) rewritten as
\begin{equation}
\label{system_appb_} \frac{\partial \theta }{\partial z}
-\frac{a(\psi_0)+b(\psi_0 )}{1+za'(\psi_0)\psi_0'(\zeta)}
\frac{\partial \theta }{\partial \zeta} =0
\end{equation}
using the change of variables
\begin{equation}
z=\eta,\quad \zeta=\tau-\eta a(\psi_0)
\end{equation}
and denoting $a'(\psi_0)\equiv da/d\psi_0$, $\psi_0'(\zeta)\equiv
d\psi_0/d\zeta$. The method of characteristics allows one to solve
Eq. (\ref{system_appb_}):
\begin{equation}
\theta(\eta,\tau)=\theta(0,\xi)\equiv\theta_0(\xi),\quad \xi\equiv
\zeta(0)
\end{equation}
with $z(s)=s$ and
\begin{equation}
\frac{d\zeta}{ds}=-\frac{a(\psi_0)+b(\psi_0)}{1+sa'(\psi_0)\psi_0'(\zeta)}.
\end{equation}
The solution of the latter equation reads
\begin{eqnarray}
s&=&\int_{\zeta}^{\xi}\frac{d\zeta'}{a[\psi_0(\zeta')]+b[\psi_0(\zeta')]}\nonumber\\
&&\times\exp\left[\int_{\psi_0(\zeta)}^{\psi_0(\zeta')}d\psi_0'
\frac{a'(\psi'_0)}{a(\psi'_0)+b(\psi_0')}\right],
\end{eqnarray}
which gives Eq. (\ref{Def_xi}).

\end{appendix}

\end{document}